**Development and validation of computable Phenotype to Identify and Characterize Kidney Health in Adult Hospitalized Patients**

Tezcan Ozrazgat-Baslanti, PhD[1], Amir Motaei, PhD[1], Rubab Islam, BS[1], Haleh Hashemighouchani, MD[1], Matthew Ruppert, BS[1], R. W. M. A. Madushani, PhD[1], Mark S. Segal, MD PhD[1,] Gloria Lipori, MBA[3], Azra Bihorac, MD MS[1], Charles Hobson MD[2]

[1]Department of Medicine, College of Medicine, University of Florida, Gainesville, FL, USA

[2]Department of Health Services Research, Management and Policy, College of Public Health and Health Professions, University of Florida, Gainesville, FL, USA

[3]University of Florida Health, Gainesville, FL, USA

Correspondence to: Azra Bihorac MD MS, Department of Medicine, Precision and Intelligent Systems in Medicine (Prisma[P]), Division of Nephrology, Hypertension, and Renal Transplantation, PO Box 100224, Gainesville, FL 32610-0254. Telephone: (352) 294-8580; Fax: (352) 392-5465; Email: abihorac@ufl.edu

Reprints will not be available from the author(s).

Words:

Short title: Computable phenotype for kidney health

Conflict of Interest Disclosures: None reported.

Funding/Support: Conflicts of Interest and Source of Funding: A.B. and T.O.B. were supported by R01 GM110240 from the National Institute of General Medical Sciences. A.B., T.O.B., and M.S. were supported by Sepsis and Critical Illness Research Center Award P50 GM-111152 from the National Institute of General Medical Sciences. A.B. was supported by a W. Martin

Smith Interdisciplinary Patient Quality and Safety Award (IPQSA). T.O.B. has received grant from Clinical and Translational Science Institute (97071) and Gatorade Trust, University of Florida. R.I. was supported by the University of Florida Medical Student Summer Research fellowship. This work was supported in part by the NIH/NCATS Clinical and Translational Sciences Award to the University of Florida UL1 TR000064. The content is solely the responsibility of the authors and does not necessarily represent the official views of the National Institutes of Health. The funders had no role in study design, data collection and analysis, decision to publish, or preparation of the manuscript. A.B., and T.O.B. had full access to all of the data in the study and take responsibility for the integrity of the data and the accuracy of the data analysis.

Previous Presentation: Partial results from this research was presented at the University of Florida Research Day.

**ABSTRACT**

**Background:** Acute kidney injury (AKI) is one of the most common complications among hospitalized patients and is central to the subsequent development of chronic kidney disease (CKD) and increased mortality. It is associated with up to five-fold increases in risk for both other serious complications and hospital death, and an increase in hospital cost of up to $28,000 per hospitalization. Timely detection of AKI and progression of AKI could avoid further injurious practices, and increase the chance for offering more effective preventive or therapeutic measures. Objective of this study is to develop and validate an electronic phenotype to identify patients with CKD and AKI.

**Methods:** Using the University of Florida Health (UFH) Integrated Data Repository as Honest Broker, we created a database with electronic health records data from a retrospective study cohort of 84,352 adult patients hospitalized at UF Health between 1/1/2012 and 4/1/2016. This repository includes demographic information, comorbidities, vital signs, laboratory values, medications with date and timestamps, and diagnoses and procedure codes for all index admission encounters as well as encounters within 12 months prior to index admission and 12 months follow-up. We developed algorithms to identify CKD and AKI based on the Kidney Disease: Improving Global Outcomes (KDIGO) criteria. We identified presence and stage of AKI by running algorithms each time a new creatinine measurement was detected. To measure diagnostic performance of the algorithms, the clinical adjudication of AKI and CKD on 300 selected cases was performed.by clinician experts with access to patient medical charts. Sensitivity, specificity, positive predictive value (PPV), and negative predictive value (NPV) of the CKD and AKI labels produced by the algorithm were compared to clinical diagnosis.

**Results:** Among 149,136 encounters, 12% had CKD by medical history, which is based on ICD-9/10 codes. Using creatinine criteria, percent of patients with CKD identified increased to 16%. Among 130,081 encounters who had sufficient data for AKI phenotyping after excluding those with end-stage renal disease on admission, AKI during hospitalization was identified in 21% of encounters. The comparison of CKD phenotyping algorithm to manual chart review performed in 300 cases yielded PPV of 0.87 (95% confidence interval (CI) 0.81-0.92), NPV of 0.99 (95% CI 0.96-1.00), sensitivity of 0.99 (95% CI 0.96-1.00), and specificity of 0.89 (95% 0.83-0.93). The comparison of AKI phenotyping algorithm to manual chart review yielded PPV of 0.99 (95% CI 0.96-1.00), NPV of 0.95 (95% CI 0.89-0.98), sensitivity 0.98 (95% CI 0.94-0.99), and specificity 0.98 (95% CI 0.93-1.00). Instead of phenotyping algorithms, if only ICD-9/10 codes were used to identify AKI, comparison to manual chart review yielded PPV of 0.78 (95% CI 0.70-0.85) and NPV of 0.40 (95% CI 0.33-0.48), specificity of 0.71 (95% CI 0.61-0.80), and sensitivity of 0.49 (95% CI 0.42-0.56). This sensitivity is very poor compared to the sensitivity of the algorithm, which also uses creatinine criteria to identify AKI.

**Conclusions:** We developed phenotyping algorithms that yielded very good performance in identification of patients with CKD and AKI in validation cohort. This tool may be useful in identifying patients with kidney disease in a large population, in assessing the quality and value of care provided to such patients and in clinical decision support tools to help providers care for these patients.

## INTRODUCTION

The advent of the electronic health record (EHR) with the availability of clinical data in digital form has transformed both clinical care and our ability to analyze that care.[1] Clinical data is the record of clinical care provided to a patient, and chart review of clinical data remains the gold standard for adjudication of clinical research questions. Claims data, based on the use of administrative codes from billable interactions between the patient and a healthcare payer, provides a global view of the patient, since virtually every healthcare interaction generates a bill. Organizations are developing enterprise data warehouses, sometimes known as integrated data repositories (IDR), which aggregate data from throughout the organization including both clinical data from the EHR and all electronic claims data.[2] Combining the broad picture of a patient seen using claims data, with the accurate clinical information contained within the EHR, could give researchers a more comprehensive view of patient care than is currently available.

Electronic phenotype is a tool to identify and characterize clinical conditions with an automated query of digital clinical record.[3] An electronic phenotype uses a defined set of data elements and logical expressions to accurately identify patients with a set of observable traits from the data contained within a data repository. These queries can be designed to identify patients with a particular condition and, ideally, to clinically stage those conditions to support observational and interventional research.[4,5] Techniques used include a combination of data mining, natural language processing techniques and regression analyses.[6]

Kidney diseases provide a good example of the challenges inherent in analyzing clinical disease data in digital clincal records, and an example of diseases to which electronic phenotyping could be used to good effect.[7,8] A query of an EHR or IDR using administrative codes alone is very poor at identifying the incidence of acute kidney injury (AKI), or the prevalence of chronic kidney disease (CKD), within any cohort of patients.[9-11] Furthermore, administrative codes do not reflect any of the disease severity metrics that are important in modern consensus definitions for AKI and CKD.[8,12] Consensus definitions of both diseases, including clinical staging, have recently been outlined by the Kidney Disease: Improving Global Outcomes (KDIGO) consortium.[13,14] Furthermore since both AKI and CKD can vary with time and interact with each other, recent epidemiologic and outcomes studies are beginning to reflect that reality. A recent publication by an Acute Disease Quality Initiative (ADQI) Workgroup proposes a nomenclature for kidney diseases across the spectrum from AKI to CKD.[15] An episode of AKI which resolves completely within 48 hours is termed “rapid reversal” AKI and is believed to have minimal clinical consequences. An episode of AKI which persists beyond 48 hours is described as “persistent” AKI, while an episode of AKI which results in renal dysfunction that persists beyond 7 days is described using the new term “Acute Kidney Disease” (AKD). Renal dysfunction persisting 90 days or longer is CKD, and CKD resulting in a need for renal replacement therapy (RRT) is end-stage kidney disease (ESKD). Patients with ESKD can undergo renal transplantation that leaves them with normal renal function or with less-severe CKD that does not require RRT.

No single electronic phenotype has been developed to identify and fully characterize both AKI and CKD within the EHR, although a preliminary and partial CKD phenotype exists[16] and methodologies have been developed to screen for AKI within the EHR or in electronic clinical registries.[17-19] Here we describe the development and validation of an automated algorithm for comprehensive identification and characterization of kidney health in electronic health records, usable both retrospectively and in real time. To the best of our knowledge this is the first electronic phenotype that combines disparate sources of EHR data to identify stages, duration and renal recovery of both acute and chronic kidney disease.

## METHODS

### Data and Participants

Using the University of Florida Health (UFH) Integrated Data Repository as Honest Broker, we have created two single-center longitudinal cohorts that integrated the comprehensive vendor-based inpatient and outpatient EHR (Epic Systems, Verona, Wisconsin, USA) available since 2011 with administrative, laboratory and claims data prior to 2011 and public datasets.[20] The Algorithm Development Cohort used the *Declare* dataset of 51,457 patients admitted to UFH between January 2000 and November 2010 (Figure 1). [21-24]

**Figure 1. Clinical Datasets**

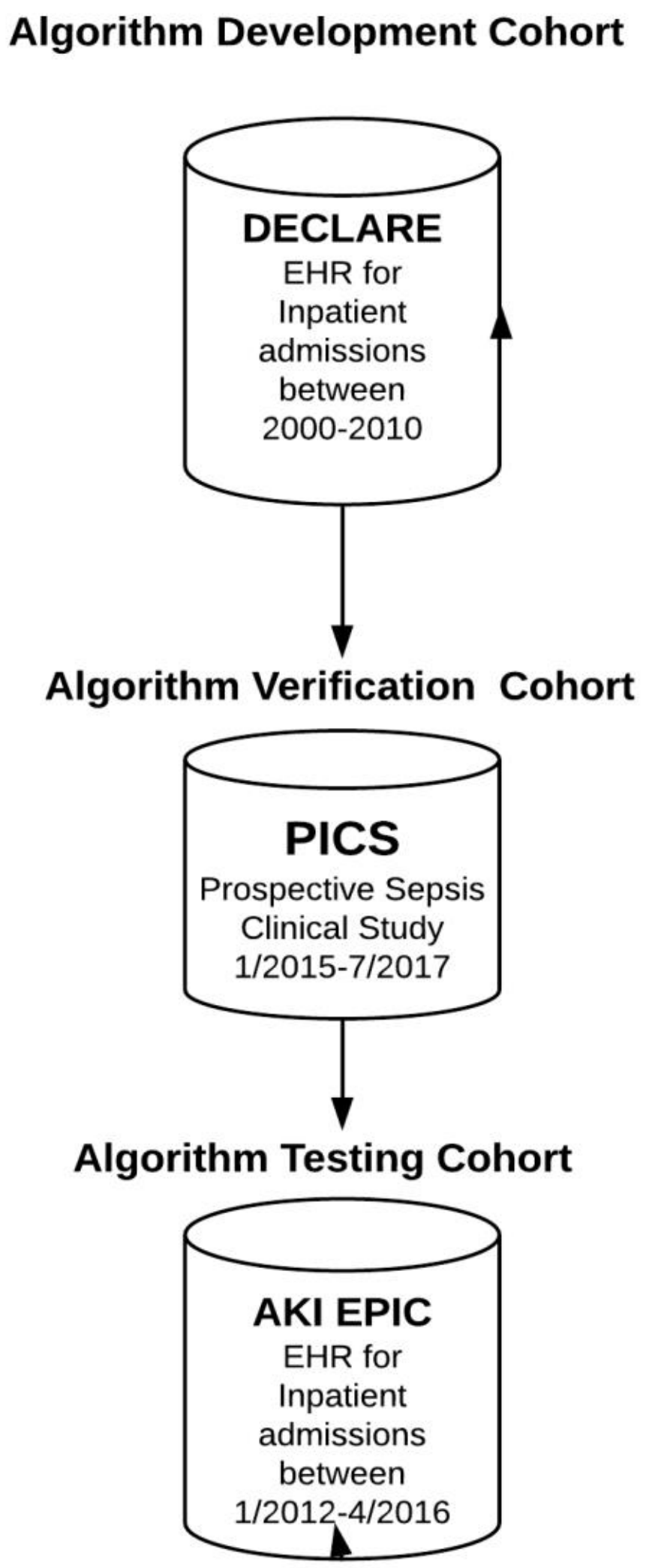

The Algorithm Verification Cohort used the *PICS* cohort that combines EHR and research data for 245 sepsis patients prospectively enrolled in longitudical cohort study at UF between January 2015 and July 2017.[25,26] The Algorithm Validation Cohort used the *AKI EPIC* dataset of 84,350 patients admitted to UFH between January 2012 and April 2016.[27] Participants in all cohorts were adults with age ≥ 18 years at the time of admission.

With the exception of PICS cohort, dataset represented the population receiving routine clinical care at the study institution. Every admission to the hospital corresponded to unique inpatient encounter. Each dataset included all diagnosis and procedure codes, structured and unstructured clinical data, demographic information, vital signs, laboratory values and medication data for all index inpatient encounters. For each of the patient in the dataset we obtained all prior administrative codes available in UFH IDR using the International Classification of Diseases, 9th and 10th revision, clinical modification (ICD-9-CM and ICD-10-CM, respectively) codes. The appropriate mapping was performed for both AKI and CKD to ensure that all appropriate codes were ascertained regardless of whether ICD-9 or ICD-10 codes are used in the EHR (Supplemental Tables 1-5). For all datasets laboratory data was available for within 12 months prior to index admission and 12 months follow-up. Data elements identified for analysis are included in the supplemental materials (Supplemental Tables 6 and 7).

**Development of the Algorithm**

We used Kidney Disease: Improving Global Outcomes (KDIGO) Clinical Practice Guideline definitions for AKI and CKD and consensus report of the Acute Disease Quality Initiative (ADQI) 16 Workgroup on renal recovery as conceptual framework for the development of algorithm for comprehensive assessment of kidney health during inpatient hospitalization (*eKidneyHealth*).[15,28,29] Using a comprehensive set of EHR data in the *Algorithm Development Cohort*, including diagnostic and procedure codes, laboratory results, clinical data and demographic information, we combined a rule-based methodology to develop algorithms and regression analysis previously proven to be replicable with strong predictive values across other institutions.[30]

The *eKidneyHealth* algorithm is a complex algorithm that unifies six rule-based flow diagrams and regression analysis in order to identify and characterize kidney health in any inpatient encounter, either in real-time or in the retrospective datasets (Figure 2).

**Figure 2. Master Flow**

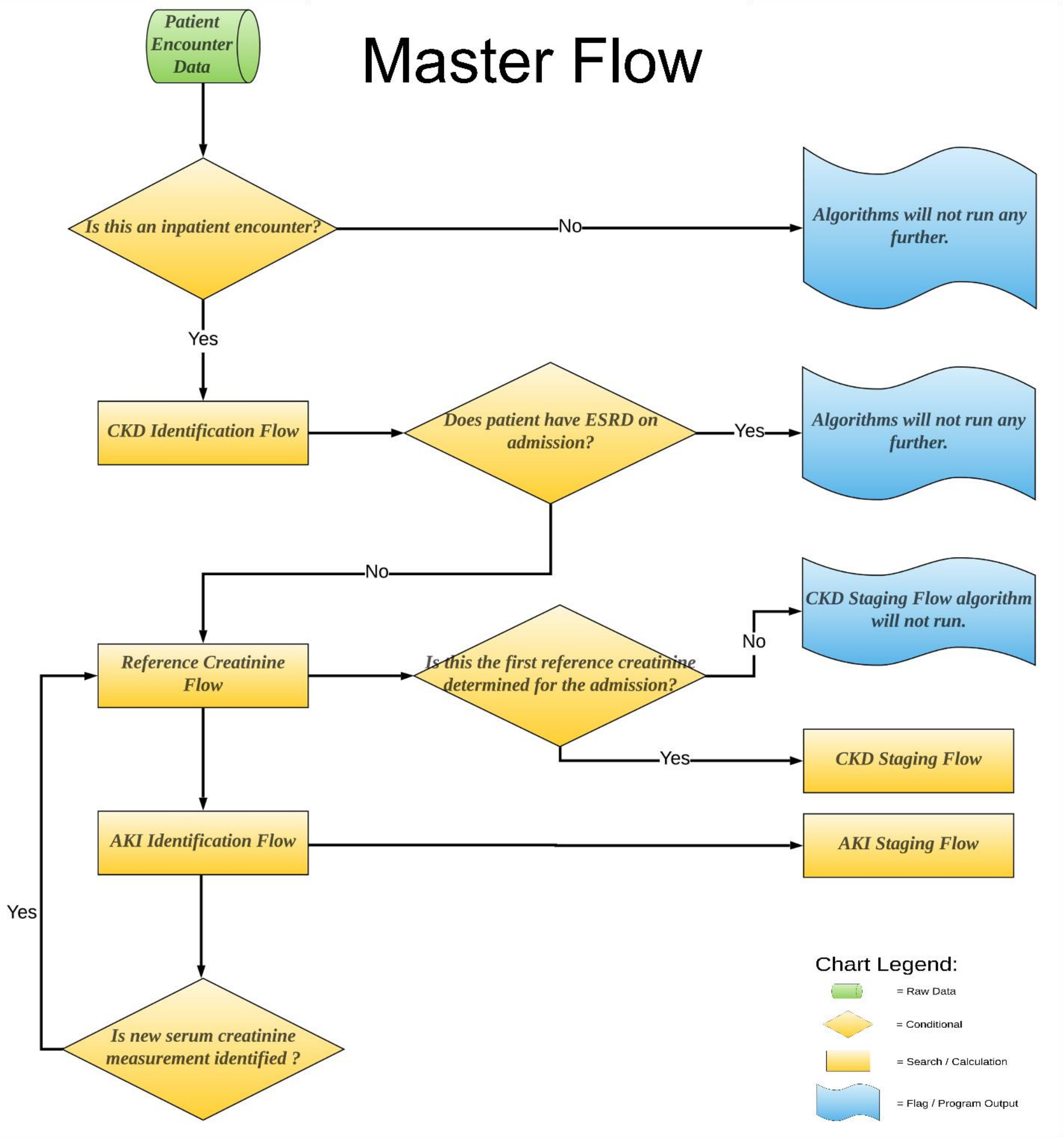

For every new inpatient encounter, immediatlly upon admission, the algorithm first determines the presence of CKD using all existing historical data prior to index admission (Figure 3).

**Figure 3. CKD Identification Flow**

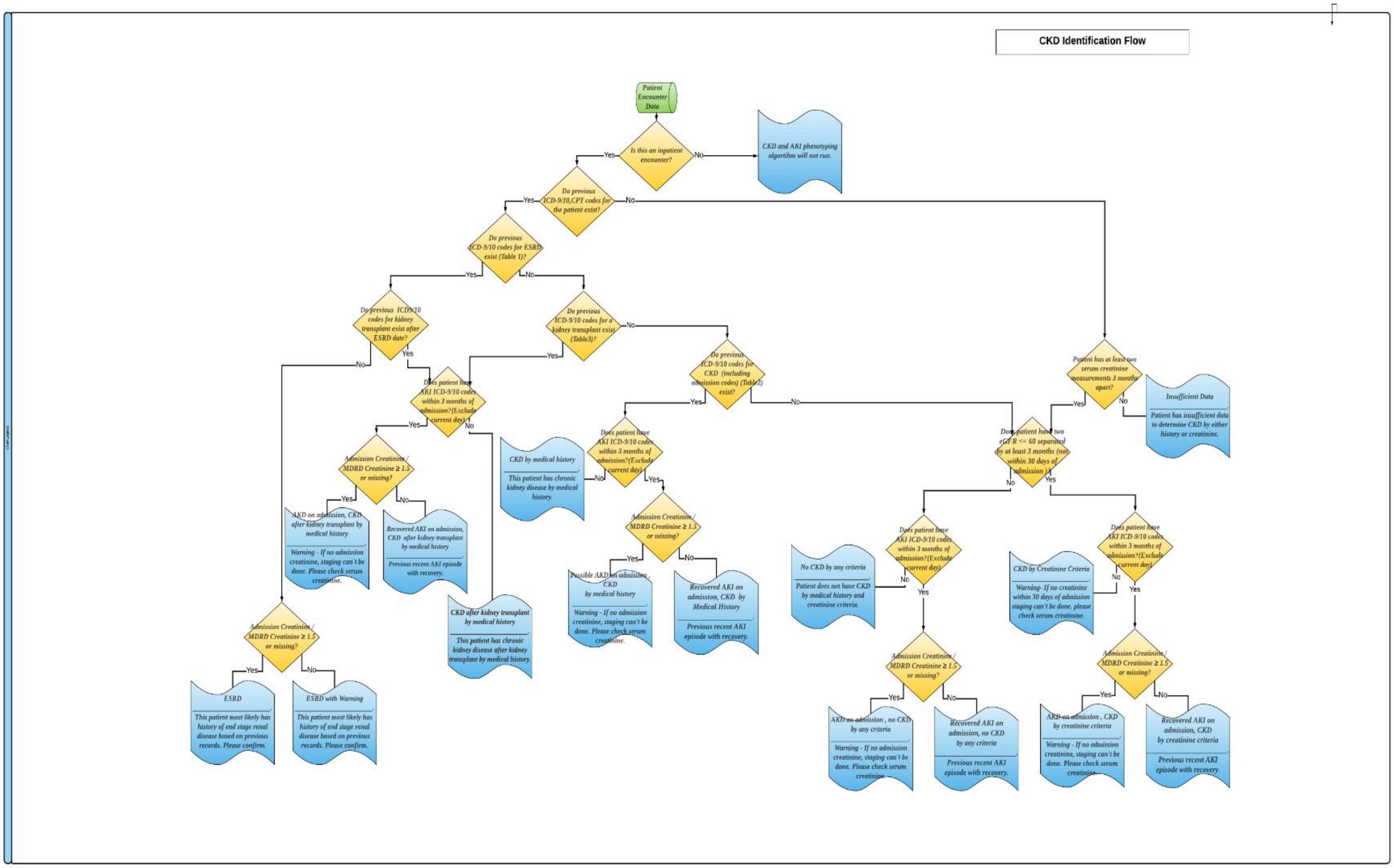

The first creatinine value during index admission triggers the assessment of reference creatinine (Figure 4) and staging of CKD (Figure 5) while any subsequent new creatinine value initiates the assessment of AKI stages, renal recovery and dynamic GFR (Figure 6).

**Figure 4. Determiantion of Reference Creatinine Flow**

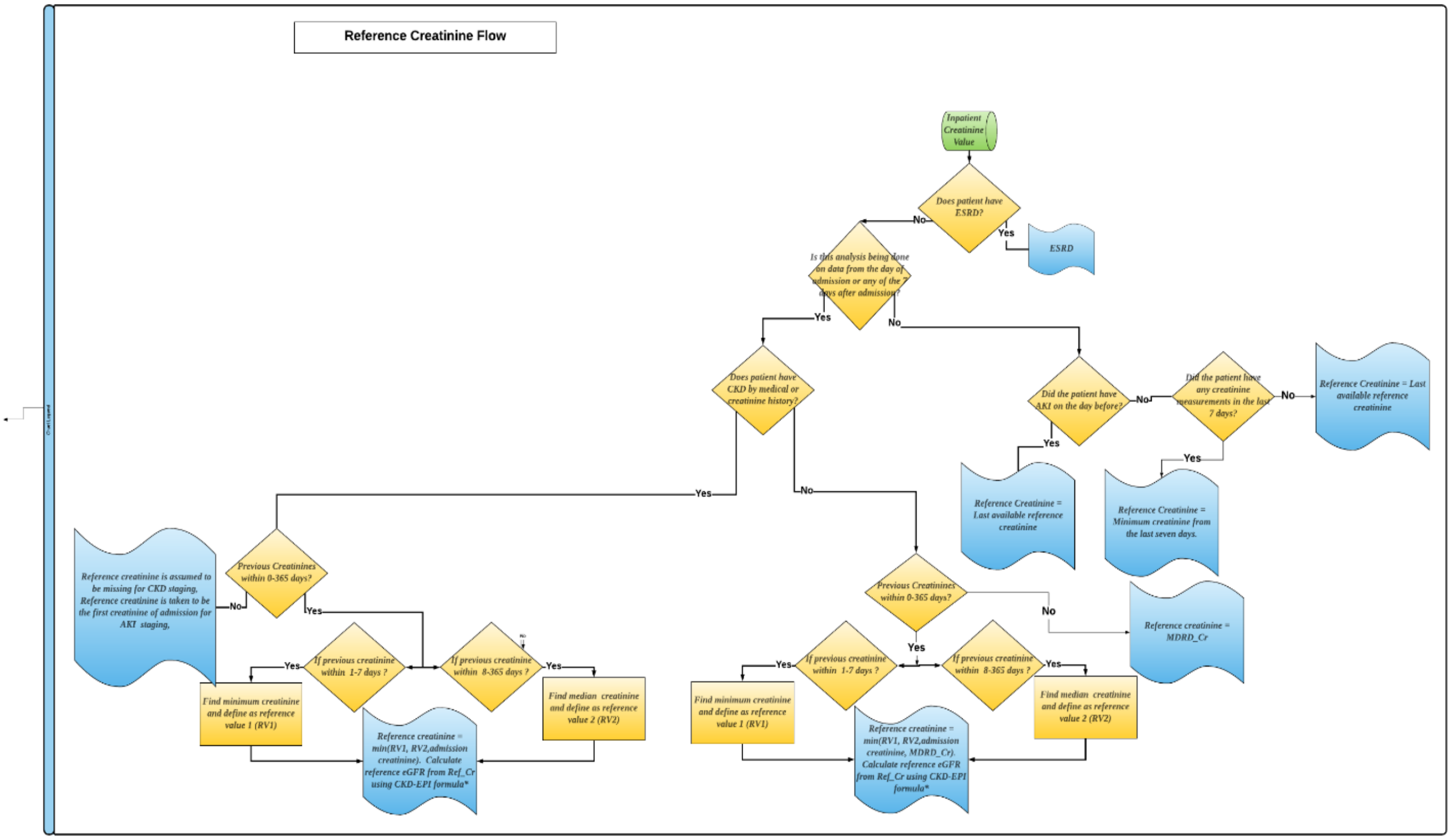

## Figure 5. CKD Staging Flow

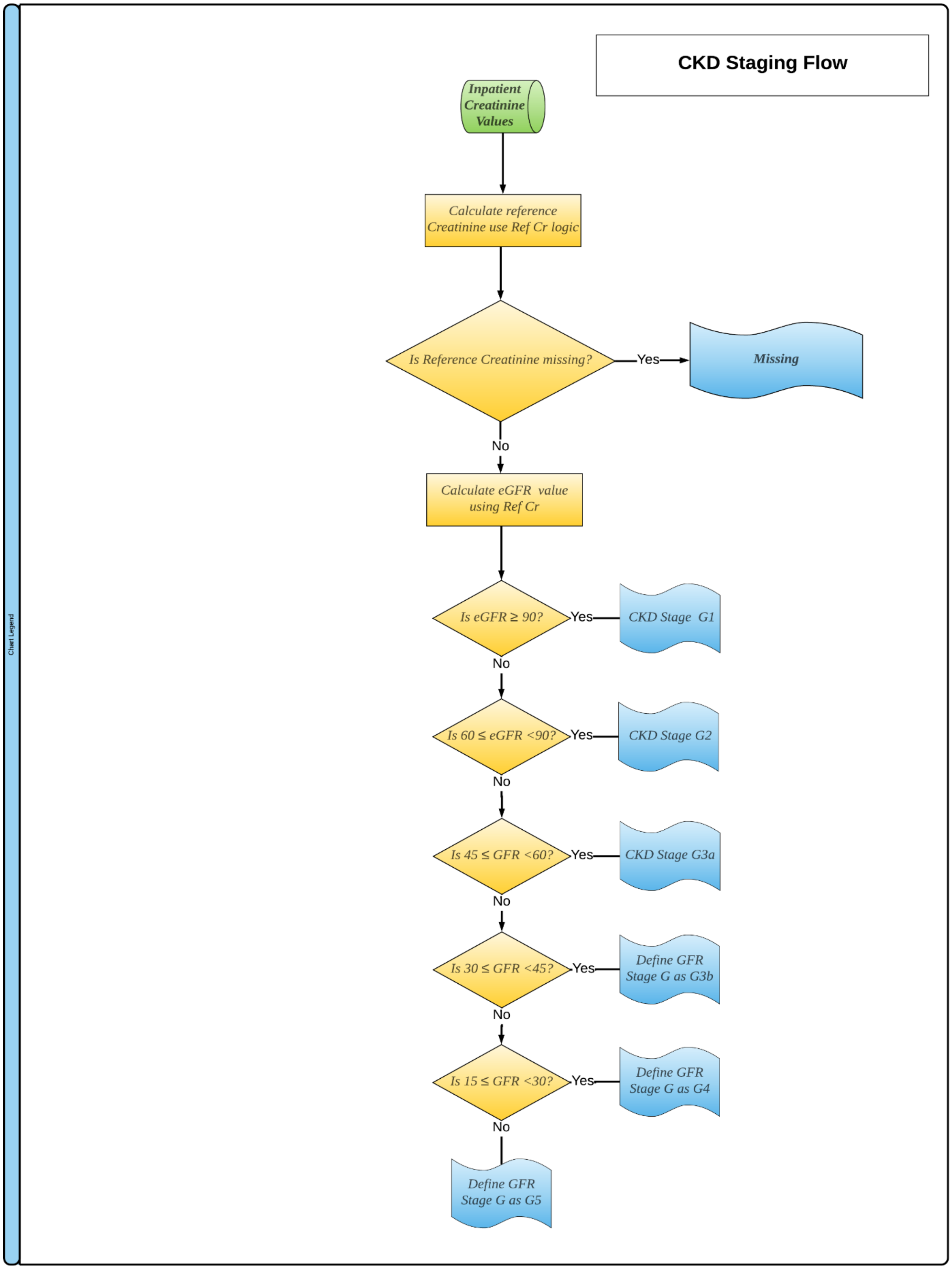

**Figure 6. AKI Identification Flow**

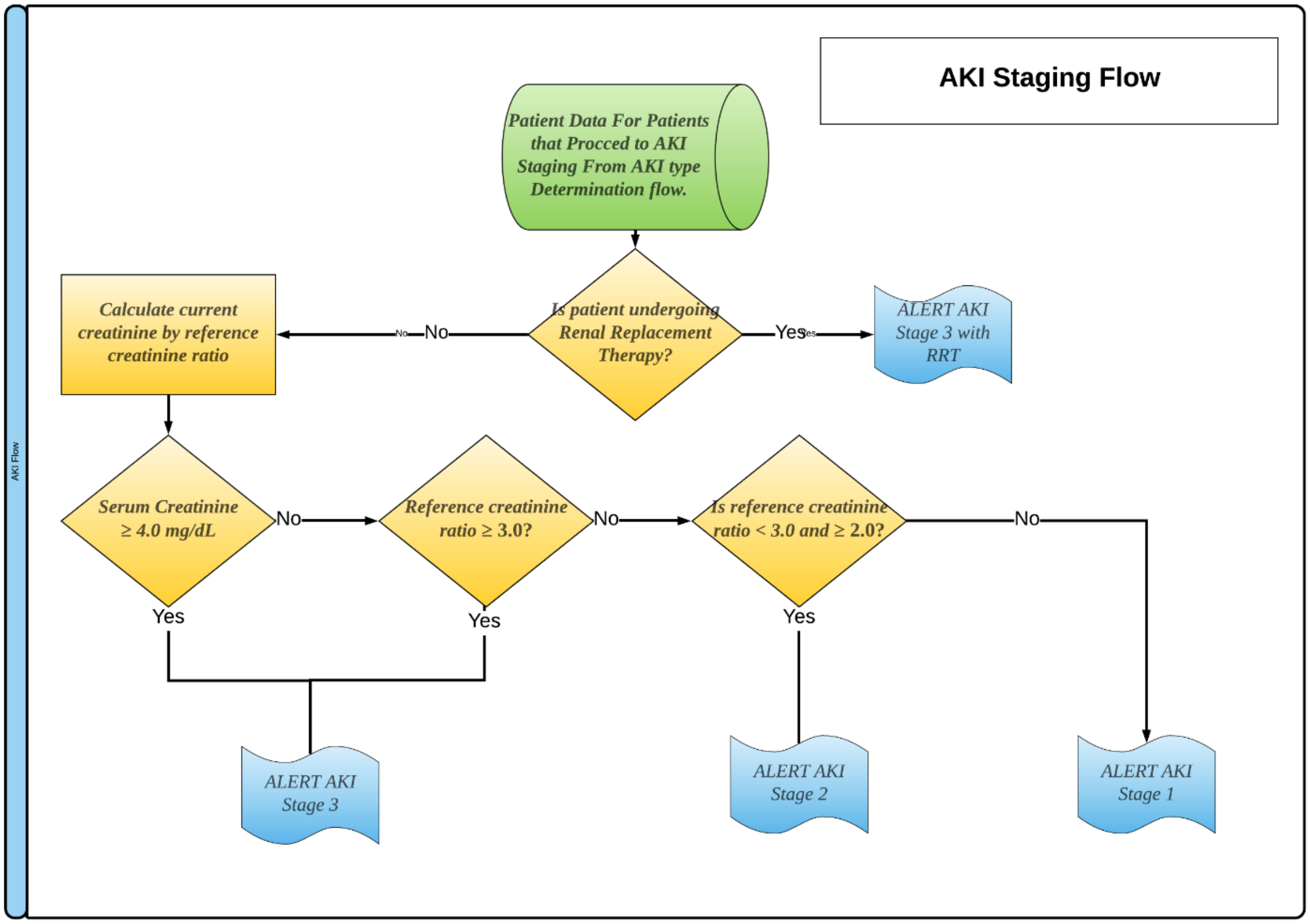

Common to both the CKD and AKI algorithms is a single "Reference Creatinine" algorithm used to identify and calculate, using available data, every patient's reference creatinine. It represents a patient's creatinine in the steady state, before any new AKI occurs, and/or after the patient recovers from any pre-existing AKI. The flows are built from data inputs, conditional decision points, search and/or calculations needed, and outputs. To identify and stage CKD for each inpatient encounter the CKD Identification, Reference Creatinine and CKD Staging algorithms are run once only using data from the day of admission and historical data. To identify and stage any new AKI during the inpatient encounter the Reference Creatinine, AKI Identification and AKI Staging algorithms are run once using data from the day of admission and historical data, and then these three algorithms are run again every time a new creatinine measurement is detected.This makes sure that any AKI that develops during an inpatient encounter is captured, and insures that the correct reference creatinine is used in doing so.

The algorithm was written in both Python 2.7 [13] and SciPy 1.0.0 [31] software. Data elements were extracted from the dataset, transformed and cleaned as needed, and then organized into Python data structures for fast query of a patients' history. Queries were developed and unit-tested for each object, and then combined into Python routines to implement the logic of the algorithms. Each patient record in the algorithm verification *PICS* dataset of 245 sepsis patients contains clinical adjudication for CKD, AKI and renal recovery status performed by two independent nehrologists as a part of a study protocol.[26] Further development and editing of the code were done using this database using clinical adjusication as the gold standard. Code subroutines were revised in an iterative manner by running the subroutines on the development and then verification cohorts, with the results compared to previous clinical adjudication.

**Identification and staging of CKD**

The CKD Identification Flow is used to determine if the patient has any evidence of CKD or ESKD, and to distinguish between patients with pre-existing CKD and those with new onset AKI (Figure 4). The algorithm first uses all available administrative codes in patient's medical record to identify patients with CKD, ESKD and any history of kidney transplantation using previously validated combination of ICD-9 or ICD-10 codes (Supplemental Tables 1-5). Patients with previous renal transplant were considered as a separate category. For subset of patients with available serum creatinine measurements prior to index hospitalization we used combination of KDIGO diagnostic criteria for CKD stage >= 3 only (risk factors for CKD and two creatinine measurements separated by at least three months with corresponding eGFR < 60 ml/min/1.72m$^2$). A subset of patients with no available previous medical history in UF EHR/IDR were marked as having insufficient data to determine CKD status. The algorithm also encounters for the episodes of AKI without renal recovery that occurred within 3 months of index admission. Outputs and definitions from the algorithm are listed in Supplemental Table 3. The CKD Staging Flow is used to calculate the patients' eGFR G-stage using the CKD-EPI formula. The reference creatinine level from the Reference Creatinine Flow is used as the standardized serum creatinine in that formula.

**Determination of Reference Creatinine**

The Reference Creatinine Flow is used to calculate a reference serum creatinine level for the patient, that is subsequently used for calculation of reference eGFR, CKD staging and AKI identification and staging. Patients with ESKD are excluded form this algorithm. The algorithm is triggered to run by every creatinine measurement identified in the inpatient encounter. Next the algorithm determines if the creatinine measurement that has triggered this run of the algorithm was obtained within the first seven days of the admission (if this run of the algorithm is for CKD this will always be true, if this run of the algorithm is for AKI the algorithm determines the date of the triggering creatinine measurement). If the index creatinine measurement is from 8 or more days after admission the algorithm identifies either the last available reference creatinine or the minimum creatinine from the previous seven days as the reference creatinine. If the index creatinine measurement is from the first seven days of the admission, a list of all serum creatinine levels with time and date stamps is used to calculate the reference creatinine. If there were previous creatinine measurements in the interval 0-7 days before admission we used the minimum creatinine level during that interval as reference value 1. If there were previous creatinine measurements in the interval 8-365 days before admission, we used the median creatinine level during that interval as reference value 2 (ref to NHS study and to Holmes paper). The reference creatinine is then the minimum of (reference value 1, reference value 2 and the admission creatinine) (Figure 4).

**Identification and staging of AKI and renal recovery**

The AKI Identification Flow is used determine if the patient has any evidence of current AKI by KDIGO criteria and to identify patients with AKI stratified by AKI maximum stages, duration if AKI (rapidly reversed AKI – duration < 48 hours, persistent AKI -duration > 48 hours and AKD – duration > 7 days) and renal recovery status (recovered vs non recovered AKI). This algorithm is triggered to run by every new measurements of serum creatinine in an inpatient encounter. We defined an episode of AKI as beginning when this algorithm identifies AKI and ending if there are two consecutive days without AKI identified, thus allowing us to identify a new episode of AKI in a patient who has recovered from a previous episode of AKI. Renal replacement status was determined daily, based on CPT codes and data elements from EHR flowsheet with order details and fluid managent for hemodialysis, peritoneal dialysis, and continuous renal replacement therapies (Supplemental tables 5-7). Since we have already excluded patients with ESKD, if the patient is undergoing RRT, identified by CPT code or by clinical evidence in the EHR, we assumed that the patient is currently being treated for AKI. We then used output from the CKD Identification algorithm, plus the date and time stamps on each serum creatinine measurement, to determine if the patient has new or persistent AKI by KDIGO criteria. Finally we determined if the patient has AKD by ADQI criteria, as well as the recovery trajectory since any previous episodes of AKI or AKD (Figure 6). With the AKI Staging Flow we determined the KDIGO AKI stage for all patients identified with AKI algorithm. If the patient was undergoing RRT the AKI is “Stage 3 with RRT”. If the patient was not undergoing RRT the current reference creatinine is used to stage the AKI (Figure 7).

**Figure 7. AKI Staging Flow**

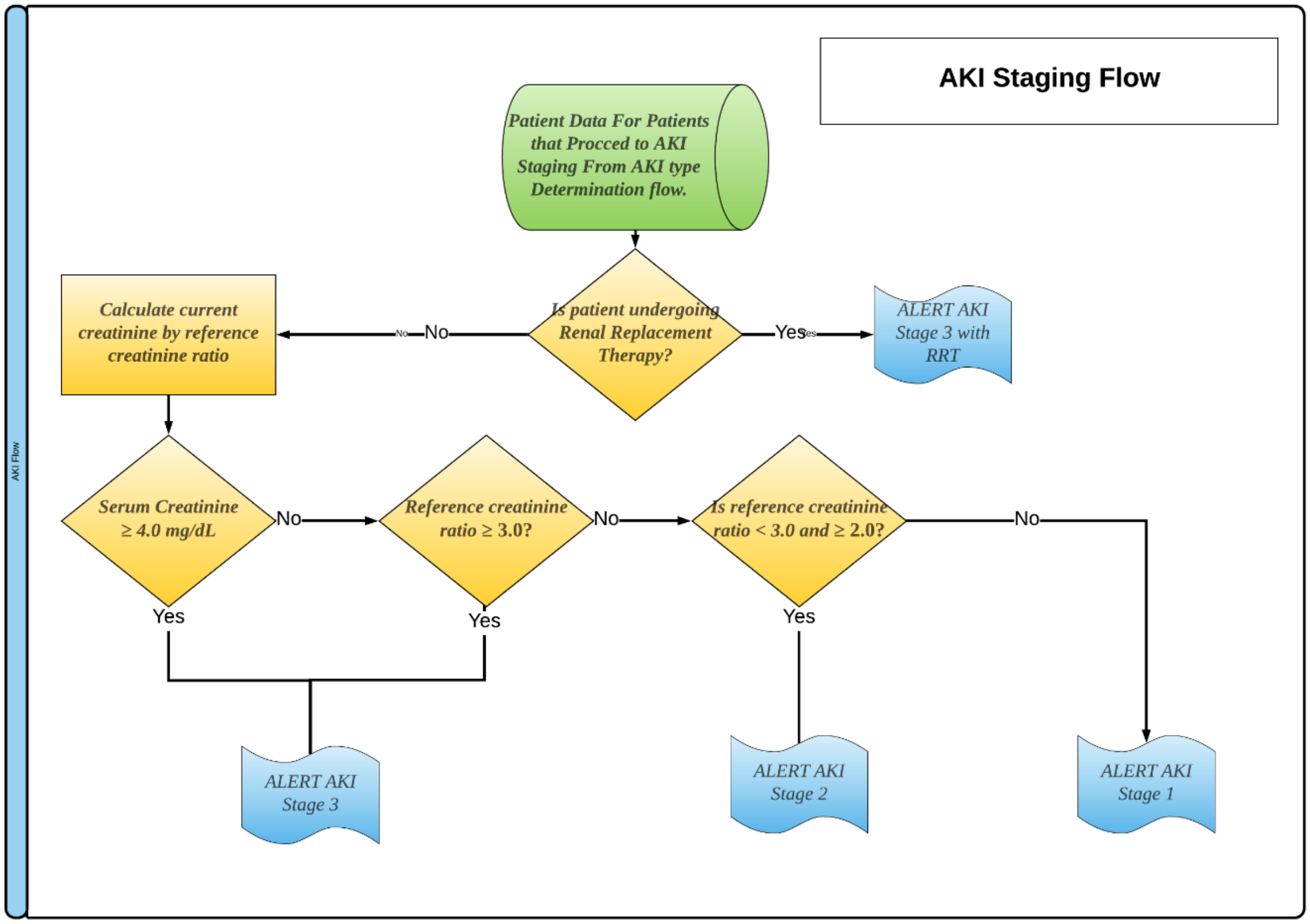

## Phenotype Algorithm Validation

The algorithms were tested and validated by comparing the perfomance of the phenotype in identifying patients with AKI and CKD to identification of these conditions by clinical experts doing review of the EHR. Performance in identifying patients with AKI was further evaluated by comparing identification by phenotype to identification by the AKI Code Algorithm.

The ability of clinician experts, with access to all patient data within a medical chart, to identify patients with disease is widely accepted as the gold standard. We enlisted three physicans (a nephrologist, an internist and a surgeon) and a medical student trained in the clinical consensus definitions of AKI and CKD to independently review the validation cohort of patients, using their EHR, to determine if the patients had CKD at the time of admission and/or AKI that developed during the hospitalization. Reviewers used physician notes, nursing notes and laboratory results to search for clinical evidence of CKD and/or AKI. Any differences in ascertainment of either CKD or AKI were arbitrated in discussion between the three reviewers while reviewing the EHR for the patient. We used the AKI Code Algorithm on the same validation cohort of patients to identify those patients who developed AKI during the hospitalization using administrative codes only. We then compared the ability of the full phenotype to identify CKD and/or AKI compared to the clinicians, and compared the ability of the full phenotype to identify AKI compared to the AKI Code Algortihm. Sensitivity, specificity, positive and negative predictive values and overall accuracy (the proportion of true classification including true positives plus true negatives) for ascertainment by the phenotype compared to ascertainment by the reference techniques were calculated with exact binomial confidence intervals. We also then reviewed the EHRs of the patients where there was a mismatch between identification by phenotype compared to identification by clinician, as an indication of how the phenotype failed and might be improved in the future.

The review sample for the phenotype algorithm validation was created by randomly selecting inpatient encounters from the *AKI Epic* database based on CKD status while stratifying each group for reference creatinine, AKI status and renal recovery. To determine how many charts would be needed for review in each subgroup we assumed that, due to the selection criteria, the expected incidence of AKI and CKD in the validation cohort would be 50% and thus a p0 for power calculation would be 0.5. We specified an odds ratio of 2.0, an $\alpha$ of 0.05 with a 2-sided test and a $\beta$ of 0.8. The power calculation, given these criteria, indicated a necessary sample size of 137 for both cases and controls or a total of 274 charts. Thus in each subgroup the patient encounters with the 25 highest reference creatinines and the 25 lowest reference creatinines were chosen for review, for a total of 300 charts. The medical record numbers for those patient encounters were pulled and used to review each patient’s record in the Epic EHR.

Statistical analyses were performed with SAS (version 9.4; SAS Institute, Inc, Cary, NC) and R software.

### Data collection and analysis

The eKidney algorithm was ran to analyze all adult inpatient encounters in the *AKI Epic* validation database to identify and characterize all AKI and CKD episodes n those encounters. If a patient had more than one episode of AKI during an inpatient encounter only the highest stage of AKI was counted. A total of 7,646,962 inpatient encounters in the *AKI Epic* database, for patients aged 18 years or greater and admitted to the UFH between January 1, 2012 and April 1, 2016, was used to construct the patient cohort for running and testing the final algorithm. We excluded all encounters that were not coded as “inpatient” or “observation” and that did not occur at either the Gainesville or Jacksonville UF campuses, all encounters with missing datetime stamps, all encounters from patients admitted in 2011 (to make sure that there was sufficient medical history in the EHR), all encounters with administrative codes for ESKD and all encounters that had no creatinine values recorded. The final cohort included 130,081 unique encounters from 49,520 patients (Figure 1).

## RESULTS

### Clinical Characteristics

The basic clinical charactristics for patients without ESKD was assessed for all 3 cohorts (Table 1). In the full cohort of 49,522 patients, approximately 51% of the population were male, and the mean age was 56 years old. The percentage of AKI and CKD was both higher among the verification cohort.

**Table 1.** Clinical characteristic for patients without ESKD for each cohort

| **Variables** | DECLARE | PICS | AKI EPIC |
|---|---|---|---|
| | (n=49,522) | (n=239) | (n=71,127[a]) |
| Age (years), mean (SD) | 56 (17) | 59 (15) | 56 (19) |
| Female sex, n (%) | 24,307 (49) | 109 (46) | 36,681 (52) |
| African-American ethnicity, n (%) | 5,923 (12) | 19 (8) | 12,112 (17) |
| Chronic kidney disease, n (%) | 4,024 (8) | 36 (15) | 5,540 (8) |
| Acute kidney injury, n (%) | 20,225 (40) | 148 (62) | 13,670 (19) |

[a] Characteristic for first encounter of each patient is used for the non-ESKD patients with sufficient data to complete AKI phenotyping.

## Distribution of chronic kidney disease groups

Among 149,136 encounters of validation cohort, 12.4% had CKD by Medical history. Using creatinine criteria, percent of patients with CKD identified increased to 16.9% (Table 2).

**Table 2.** Distribution of chronic kidney disease groups

| | **AKI EPIC Cohort n (%)** |
|---|---|
| **Overall** | 149136 (100) |
| Insufficient Data ( No CKD with warning) | 630 (0.4) |
| ***No CKD*** | 123268 (82.7) |
| No CKD by Medical History Or Creatinine Criteria and no recent AKI episode[a] | 118392 (79.4) |
| No CKD by Medical History Or Creatinine Criteria, Recovered recent AKI on Admission | 3483 (2.3) |
| No CKD by Medical History Or Creatinine Criteria, Non-recovered recent AKI (AKD) on admission | 1393 (0.9) |
| ***CKD*** | 25238 (16.9) |
| **CKD by Medical History** | 18557 (12.4) |
| CKD by Medical History and no recent AKI episode[a] | 12365 (8.3) |
| CKD by Medical History, Recovered recent AKI on Admission | 2858 (1.9) |
| CKD by Medical History, Non-recovered recent AKI (AKD) on Admission | 3334 (2.2) |
| **CKD by Creatinine Criteria** | 4937 (3.3) |
| CKD by Creatinine Criteria and no recent AKI episode[a] | 3974 (2.7) |
| CKD by Creatinine Criteria, Recovered recent AKI on Admission | 631 (0.4) |
| CKD by Creatinine Criteria, Non-recovered recent AKI (AKD) on Admission | 332 (0.2) |
| **CKD after kidney transplant** | 1744 (1.2) |
| CKD after kidney transplant and no recent AKI episode[a] | 1038 (0.7) |
| CKD after kidney transplant, Recovered recent AKI on Admission | 187 (0.1) |
| CKD after kidney transplant, Non-recovered recent AKI (AKD) on Admission | 519 (0.3) |

| CKD Stages among all encounters with CKD | |
|---|---|
| G1 (eGFR ≥ 90 ml/min/1.73m$^2$) | 4424 (17.5) |
| G2 (90>eGFR ≥ 60 ml/min/1.73m$^2$) | 8098 (32.1) |
| G3a (60>eGFR ≥ 45 ml/min/1.73m$^2$) | 5668 (22.5) |
| G3b (45>eGFR ≥ 30 ml/min/1.73m$^2$) | 4314 (17.1) |
| G4 (30>eGFR ≥ 15 ml/min/1.73m$^2$) | 2071 (8.2) |
| G5 ( eGFR < 15 ml/min/1.73m$^2$) | 410 (1.6) |
| No staging can be done | 253 (1) |

Abbreviations: AKD, acute kidney disease; CKD, chronic kidney disease; eGFR, estimated glomerular filtration rate.
[a] Recent AKI episode defined by the presence of ICD9 or 10 codes documented in EHR in the three months prior to admission

### AKI Charachtristics

Among 130,081 encounters who had sufficient data for AKI phenotyping after excluding those with end-stage renal disease on admission, AKI during hospitalization was identified in 21% of encounters (Table 3). The maximum AKI stage was mostly stage 1 (63%), while AKI stage 2 and 3 were the maximum stage identified in 19% and 18% of the patients, respectively. Twelve percent of patients developed more than one episode of AKI.

**Table 3.** Renal characteristics among encounters with no end-stage renal disease on admission

| | **AKI EPIC Cohort (N=130,081)** |
|---|---|
| **No AKI during hospitalization, n (%)** | 103089 (79.2) |
| **AKI during hospitalization, n (%)** | 26992 (20.8) |
| **Maximum AKI Stage, n (%)** | |
| Stage 1 | 16949 (62.8) |
| Stage 2 | 5236 (19.3) |
| Stage 3 (with or without RRT) | 4807 (17.8) |
| RRT, n (%) | 1306 (4.2) |
| Number of days on RRT, median (25$^{th}$, 50$^{th}$, 75$^{th}$) | 10 (5, 20) |
| Recurrent AKI, n (%) | 3310 (12.3) |
| AKI duration, days, median (25$^{th}$, 50$^{th}$, 75$^{th}$) | 2 (1, 5) |
| **AKI trajectories** | |
| Rapidly reversed AKI | 10163 (37.7) |
| Persistent AKI | 16829 (62.3) |

Abbreviations: AKI, acute kidney injury; RRT, renal replacement therapy
Recurrent AKI is defined as number of AKI episodes more than one.

### Comparison of phenotyping algorithms performance to manual chart review.

CKD and AKI phenotyping algorithms performed well with diagnostic performance measures above 0.90. (Tables 4, 5).

**Table 4.** Comparison of chronic kidney disease phenotyping algorithm performance to manual chart review.

| | Manual chart review | | |
|---|---|---|---|
| **Phenotyping Algorithm** | **Case** | **Control** | **Total** |
| **Case** | 131 | 19[a] | 150 |
| **Control** | 1[b] | 149 | 150 |
| **Total** | 132 | 168 | 300 |
| **PPV (95% CI)** | 0.87 (0.81, 0.92) | | |
| **NPV (95% CI)** | 0.99 (0.96, 1.00) | | |
| **Sensitivity (95% CI)** | 0.99 (0.96, 1.00) | | |
| **Specificity (95% CI)** | 0.89 (0.83, 0.93) | | |
| **Accuracy (95% CI)** | 0.93 (0.90, 0.96) | | |

Abbreviations. CI, confidence interval; NPV, negative predictive value; PPV, positive predictive value; ICD, International Classification of Diseases.
Reasons for mismatches between phenotyping algorithm and manual chart review includes:
[a]Assignment of wrong ICD code for patient who had AKI (n=4), Assignment of wrong ICD code (n=2), Assignment of wrong ICD code for nephrotic syndrome (n=4), and Non-specific CKD code for patient who had AKI (n=9)
[b]Algorithm missed CKD by creatinine criteria (n=1)

**Table 5.** Comparison of Acute kidney injury phenotyping algorithm performance and ICD-9/10 codes to manual chart review.

| Manual chart review | | | |
|---|---|---|---|
| **Phenotyping Algorithm** | **Case** | **Control** | **Total** |
| **Case** | 198 | 2[a] | 200 |
| **Control** | 5[b] | 95 | 100 |
| **Total** | 203 | 97 | 300 |
| **PPV (95% CI)** | 0.99 (0.96, 1.00) | | |
| **NPV (95% CI)** | 0.95 (0.89, 0.98) | | |
| **Sensitivity (95% CI)** | 0.98 (0.94, 0.99) | | |
| **Specificity (95% CI)** | 0.98 (0.93, 1.00) | | |
| **Accuracy (95% CI)** | 0.98 (0.95, 0.99) | | |

Abbreviations. CI, confidence interval; NPV, negative predictive value; PPV, positive predictive value; ICD, International Classification of Diseases.
Reasons for mismatches between phenotyping algorithm and manual chart review includes:
[a]Reference creatinine wrong based on erroneous laboratory measurement (n=2)
[b]Wrong reference creatinine due to insufficient creatinine history for CKD patient (n=2) and Wrong reference creatinine due to wrong CKD code assignment (n=3)

## DISCUSSION

Acute kidney injury and chronic kidney disease are common, interrelated, and cause significant morbidity and mortality for patients. [32,33] Outcomes are the worst for patients with severe AKI and/or CKD, but they degrade even for patients with mild and moderate disease.[34,35] The ability to automatically identify and stage patients with AKI and CKD within a data repository of clinical data would facilitate clinical and health services research and, if done in real time, could greatly facilitate clinical decision support.[22,36] In this work we developed and validated a novel methodology to identify adult patients with kidney disease using both clinical and claims data in an integrated data repository. We applied the technique of electronic phenotyping in a large retrospective cohort of hospitalized patients to translate a clinician's approach to assessing AKI and CKD into a set of computer algorithms. We developed a phenotype for adult kidney health that allowed us to identify patients with both AKI and CKI, to clinically stage those diseases and to assess any recovery from AKI that occurs during the inpatient hospital stay. The overall algorithm is constructed from three algorithms to identify CKD plus three algorithms to identify AKI, with both using a common "Reference Creatinine" algorithm used to identify every patient's baseline creatinine and which is updated every time a new serum creatinine value is identified. The outputs from the overall algorithm include CKD by medical history and/or creatinine criteria including ESKD, AKD and recovered AKI, KDIGO G-stage of any CKD, new and persistent AKI, recovery from AKI and from AKD, and KDIGO stage of any AKI.

Kidney disease is commonly divided clinically into acute kidney injury (AKI) and chronic kidney disease (CKD), although both diseases exist on a spectrum of overall kidney health.[37] While AKI is often seen as an acute complication of surgery, as drug toxicity or as a result of trauma or sepsis, CKD is a chronic illness most commonly associated with diabetes and/or hypertension. Despite this clinical and logical separation, there is significant overlap between the entities.[38] An episode of AKI can become CKD if there is no recovery after the initial insult. Patients with CKD are at higher risk for developing AKI compared to patients without CKD, while an episode of AKI puts a patient at higher risk for developing CKD even if the episode of AKI resolves completely. Both AKI and CKD increase a patient's risk for other diseases, especially cardiovascular diseases. [23,39]

Patients with clinical disease can be identified within a database of clinical information, as is contained within a database of EHRs, or they can be identified within a database billing or claims data made up of administrative codes as in the National Inpatient Sample.[40,41] Research using clinical data once required laborious and time-consuming manual review of paper charts, but can now be done quickly through automated review of EHRs. Clinical data contains the full richness of a patient's medical history, but is limited to information from only one provider, one hospital, or at most one delivery system, and care provided outside of those constraints is not available for analysis.[42] Claims data can provide a global picture of care provided to any patient, but it cannot provide a complete clinical picture as there is often minimal or no information on disease severity and coding accuracy can be problematic.[43,44] While the clinical information contained within billing data is becoming more dense and granular with every revision of the coding systems, using billing data in the analysis of clinical disease requires much more clinical information than is currently available in claims data. An automated approach to identifying and characterizing disease that combines the global picture seen in administrative codes, with the clinical detail provided in EHR data, could provide accurate and reliable inferences about the presence and severity of clinical illness.[45,46]

Originally developed by genomics researchers, to query EHRs to accurately capture all patients with rare genetic diseases, the electronic phenotype is finding wider use in both clinical and health services research applications.[47] The goal in electronic phenotyping is to accurately identify patients with a specific observable trait from the large volumes of imperfect and/or incomplete practice-based data contained within the EHR [3] The Electronic Medical Records and Genomics (eMERGE) Network, an international consortium of genetics researchers, publishes best practices in developing and using phenotyping algorithms and helped establish an online repository, the Phenotype Knowledgebase (https://phekb.org), where phenotypes can be published and then validated at other participating institutions.[30,48] Phenotypes that include clinical data such as medication records, clinical progress notes, and laboratory and radiology data notes along with administrative codes have demonstrated much improved disease identification as measured by positive and negative predictive values compared to identification using administrative data alone.[42,49]

We evaluated the ability of the kidney health phenotype to identify patients with kidney disease against the gold standard of clinical chart review and against the process of identifying patients solely by using administrative data. The performance of the phenotype in identifying AKI and CKD exceeds existing tools and greatly improves upon the ability to capture AKI using administrative codes alone.[50-52] Importantly the phenotype performs well across the spectrum of disease severity including minor stages of AKI. Recent studies have demonstrated the ability of identification tools that emphasize severe kidney injury, such as the Major Adverse Kidney Events by 30 days (MAKE30) composite of death, new renal replacement therapy, or persistent renal dysfunction to retrospectively identify AKI in the EHR with high sensitivity and specificity.[53] However methods to capture mild and moderate AKI in the EHR are lacking.[7] Mild to moderate AKI is much more common than severe AKI, and has become much more appreciated with the introduction of consensus definitions for AKI.[54] Recent work using modern consensus definitions of AKI have shown that when properly defined AKI is one of the most prevalent complications seen in hospitalized patients, is associated with other adverse events, results in an increase in resource utilization and increases in both short- and long-term mortality.[32,34,55,56] The ability to more accurately identify AKI and CKD in retrospective analysis of large medical databases will improve the ability to measure the costs of care, the costs of complications of care and the assessment of provider clinical performance.[57,58] Perhaps even more importantly, the ability to accurately identify AKI in real-time or close to real-time analysis of data within the EHR may lead to earlier and better care for the patients that have just been identified with new or worsening AKI.[50-52]

## Limitations

This work, despite the performance of the phenotype, has some significant limitations. One important limitation is related to how the the electronic phenotype was developed. To the extent that the phenotype relies on administrative codes for AKI and CKD it is dependant upon accurate and precise coding of diseases in the EHR, a reality that is rarely if ever achieved. Since identification of both AKI and CKD are dependant upon by changes in serum creatinine and in a determination of baseline creatinine, the phenotype misses other important clinical signs of kidney injury and illness. Oliguria, or decreased urine output, is an important predictor and early sign of kidney injury and is not captured in this phenotype. Changes is urine and serum biomarkers, an important and evolving technology to identify AKI and CKD, is not captured in the phenotype. Finally the phenotype as built is largely diagnosis agnostic. While it does capture administrative codes for a wide variety of kidney diseases, at the end it just identifies a patient with AKI and/or CKD with KDIGO staging, but it does not capture specific etiologies of either condition and thus misses important clinical information related to etiology. The performance of the phenotype needs to be further evaluated, to determine its ability to accurately measure the temporal aspects of AKI and CKD and renal recovery if any, and to determine its performance in the outpatient setting.

## Conclusion

We have described the development and validation of an electronic phenotype for kidney health in hospitalized adult patients. This tool could improve studies into the epidemiology of AKI and CKD, could improve the assessment of the costs of these conditions and the quality of care, and could hasten the development of effective and early alerts for AKI.

## SUPPLEMENTAL TABLES

**Table S1.** Administrative codes used for end stage kidney disease

| ICD Code | Explanation |
|---|---|
| **ICD-9-CM Diagnosis** | |
| 585.6 | End stage kidney disease |
| V45.1 | Renal dialysis status<br>Excludes:admission for dialysis treatment or session (V56.0) |
| V45.11 | Renal dialysis status<br>Hemodialysis status<br>Patient requiring intermittent renal dialysis<br>Peritoneal dialysis status<br>Presence of arterial-venous shunt (for dialysis) |
| V45.12 | Noncompliance with renal dialysis |
| **ICD-10-CM Diagnosis** | |
| N18.6 | End stage kidney disease |
| Z91.15 | Patient's noncompliance with renal dialysis |

**Table S2.** Administrative codes used for chronic kidney disease

| ICD Code | Explanation |
|---|---|
| **ICD-9-CM Diagnosis** | |
| 403.00 | Hypertensive chronic kidney disease, malignant, with chronic kidney disease stage I through stage IV, or unspecified |
| 403.01 | Hypertensive chronic kidney disease, malignant, with chronic kidney disease stage V or end stage kidney disease |
| 403.10 | Hypertensive chronic kidney disease, benign, with chronic kidney disease stage I through stage IV, or unspecified |
| 403.11 | Hypertensive chronic kidney disease, benign, with chronic kidney disease stage V or end stage kidney disease |
| 403.90 | Hypertensive chronic kidney disease, unspecified, with chronic kidney disease stage I through stage IV, or unspecified |
| 403.91 | Hypertensive chronic kidney disease, unspecified, with chronic kidney disease stage V or end stage kidney disease |
| 404.00 | Hypertensive heart and chronic kidney disease, malignant, without heart failure and with chronic kidney disease stage I through stage IV, or unspecified |
| 404.01 | Hypertensive heart and chronic kidney disease, malignant, with heart failure and with chronic kidney disease stage I through stage IV, or unspecified |
| 404.02 | Hypertensive heart and chronic kidney disease, malignant, without heart failure and with chronic kidney disease stage V or end stage kidney disease |
| 404.03 | Hypertensive heart and chronic kidney disease, malignant, with heart failure and with chronic kidney disease stage V or end stage kidney disease |
| 404.10 | Hypertensive heart and chronic kidney disease, benign, without heart failure and with chronic kidney disease stage I through stage IV, or unspecified |
| 404.11 | Hypertensive heart and chronic kidney disease, benign, with heart failure and with chronic kidney disease stage I through stage IV, or unspecified |
| 404.12 | Hypertensive heart and chronic kidney disease, benign, without heart failure and with chronic kidney disease stage V or end stage kidney disease |
| 404.13 | Hypertensive heart and chronic kidney disease, benign, with heart failure and chronic kidney disease stage V or end stage kidney disease |
| 404.90 | Hypertensive heart and chronic kidney disease, unspecified, without heart failure and with chronic kidney disease stage I through stage IV, or unspecified |
| 404.91 | Hypertensive heart and chronic kidney disease, unspecified, with heart failure and with chronic kidney disease stage I through stage IV, or unspecified |
| 404.92 | Hypertensive heart and chronic kidney disease, unspecified, without heart failure and with chronic kidney disease stage V or end stage kidney disease |
| 404.93 | Hypertensive heart and chronic kidney disease, unspecified, with heart failure and chronic kidney disease stage V or end stage kidney disease |
| 581 | Nephrotic syndrome |
| 581.0 | Nephrotic syndrome with lesion of proliferative glomerulonephritis |
| 581.1 | Nephrotic syndrome with lesion of membranous glomerulonephritis |
| 581.2 | Nephrotic syndrome with lesion of membranoproliferative glomerulonephritis |
| 581.3 | Nephrotic syndrome with lesion of minimal change glomerulonephritis |
| 581.8 | Nephrotic syndrome with other specified pathological lesion in kidney |
| 581.81 | Nephrotic syndrome in diseases classified elsewhere |
| 581.89 | Nephrotic syndrome with other specified pathological lesion in kidney |
| 581.9 | Nephrotic syndrome with unspecified pathological lesion in kidney |
| 582 | Chronic glomerulonephritis |
| 582.0 | Chronic glomerulonephritis with lesion of proliferative glomerulonephritis |
| 582.1 | Chronic glomerulonephritis with lesion of membranous glomerulonephritis |
| 582.2 | Chronic glomerulonephritis with lesion of membranoproliferative glomerulonephritis |
| 582.4 | Chronic glomerulonephritis with lesion of rapidly progressive glomerulonephritis |
| 582.8 | Chronic glomerulonephritis with other specified pathological lesion in kidney |
| 582.81 | Chronic glomerulonephritis in diseases classified elsewhere |
| 582.89 | Chronic glomerulonephritis with other specified pathological lesion in kidney |
| 582.9 | Chronic glomerulonephritis with unspecified pathological lesion in kidney |
| 583 | Nephritis and nephropathy, not specified as acute or chronic |
| 583.0 | Nephritis and nephropathy, not specified as acute or chronic, with lesion of proliferative glomerulonephritis |
| 583.1 | Nephritis and nephropathy, not specified as acute or chronic, with lesion of membranous glomerulonephritis |
| 583.2 | Nephritis and nephropathy, not specified as acute or chronic, with lesion of membranoproliferative glomerulonephritis |
| 583.4 | Nephritis and nephropathy, not specified as acute or chronic, with lesion of rapidly progressive glomerulonephritis |
| 583.6 | Nephritis and nephropathy, not specified as acute or chronic, with lesion of renal cortical necrosis |
| 583.7 | Nephritis and nephropathy, not specified as acute or chronic, with lesion of renal medullary necrosis |
| 583.8 | Nephritis and nephropathy not specified as acute or chronic with other specified pathological lesion in kidney |
| 583.81 | Nephritis and nephropathy, not specified as acute or chronic, in diseases classified elsewhere |
| 583.89 | Nephritis and nephropathy, not specified as acute or chronic, with other specified pathological lesion in kidney |
| 583.9 | Nephritis and nephropathy, not specified as acute or chronic, with unspecified pathological lesion in kidney |
| 585 | Chronic kidney disease (ckd) |
| 585.1 | Chronic kidney disease, Stage I |
| 585.2 | Chronic kidney disease, Stage II (mild) |
| 585.3 | Chronic kidney disease, Stage III (moderate) |
| 585.4 | Chronic kidney disease, Stage IV (severe) |
| 585.9 | Chronic kidney disease, unspecified |
| 586 | Renal failure, unspecified |
| 250.40 | Diabetes with renal manifestations, type II or unspecified type, not stated as uncontrolled |
| 250.41 | Diabetes with renal manifestations, type I [juvenile type], not stated as uncontrolled |
| 250.42 | Diabetes with renal manifestations, type II or unspecified type, uncontrolled |
| 250.43 | Diabetes with renal manifestations, type I [juvenile type], uncontrolled |
| 588.8 | Other specified disorders resulting from impaired renal function |
| 588.81 | Secondary hyperparathyroidism (of renal origin) |
| 588.89 | Other specified disorders resulting from impaired renal function |
| 588.9 | Unspecified disorder resulting from impaired renal function |
| 753.13 | Polycystic kidney, autosomal dominant |
| **ICD-10-CM Diagnosis** | |
| I12.0 | Hypertensive chronic kidney disease with stage 5 chronic kidney disease or end stage kidney disease |
| I12.9 | Hypertensive chronic kidney disease with stage 1 through stage 4 chronic kidney disease, or unspecified chronic kidney disease |
| I13.0 | Hypertensive heart and chronic kidney disease with heart failure and stage 1 through stage 4 chronic kidney disease, or unspecified chronic kidney disease |
| I13.1 | Hypertensive heart and chronic kidney disease without heart failure |
| I13.10 | Hypertensive heart and chronic kidney disease without heart failure with stage 1 through stage 4 chronic kidney disease, or unspecified chronic kidney disease |
| I13.11 | Hypertensive heart and chronic kidney disease without heart failure with stage 5 chronic kidney disease, or end stage kidney disease |
| I13.2 | Hypertensive heart and chronic kidney disease with heart failure and with stage 5 chronic kidney disease, or end stage kidney disease |
| N01 | Rapidly progressive nephritic syndrome |
| N01.0 | Rapidly progressive nephritic syndrome with minor glomerular abnormality |
| N01.1 | Rapidly progressive nephritic syndrome with focal and segmental glomerular lesions |
| N01.2 | Rapidly progressive nephritic syndrome with diffuse membranous glomerulonephritis |
| N01.3 | Rapidly progressive nephritic syndrome with diffuse mesangial proliferative glomerulonephritis |
| N01.4 | Rapidly progressive nephritic syndrome with diffuse endocapillary proliferative glomerulonephritis |
| N01.5 | Rapidly progressive nephritic syndrome with diffuse mesangiocapillary glomerulonephritis |
| N01.6 | Rapidly progressive nephritic syndrome with dense deposit disease |
| N01.7 | Rapidly progressive nephritic syndrome with diffuse crescentic glomerulonephritis |
| N01.8 | Rapidly progressive nephritic syndrome with other morphologic changes |
| N01.9 | Rapidly progressive nephritic syndrome with unspecified morphologic changes |
| N02.0 | Recurrent and persistent hematuria with minor glomerular abnormality |
| N02.1 | Recurrent and persistent hematuria with focal and segmental glomerular lesions |
| N02.2 | Recurrent and persistent hematuria with diffuse membranous glomerulonephritis |
| N02.3 | Recurrent and persistent hematuria with diffuse mesangial proliferative glomerulonephritis |
| N02.4 | Recurrent and persistent hematuria with diffuse endocapillary proliferative glomerulonephritis |
| N02.5 | Recurrent and persistent hematuria with diffuse mesangiocapillary glomerulonephritis |
| N02.6 | Recurrent and persistent hematuria with dense deposit disease |
| N02.7 | Recurrent and persistent hematuria with diffuse crescentic glomerulonephritis |
| N02.8 | Recurrent and persistent hematuria with other morphologic changes |
| N02.9 | Recurrent and persistent hematuria with unspecified morphologic changes |
| N03 | Chronic nephritic syndrome |
| N03.0 | Chronic nephritic syndrome with minor glomerular abnormality |
| N03.1 | Chronic nephritic syndrome with focal and segmental glomerular lesions |
| N03.2 | Chronic nephritic syndrome with diffuse membranous glomerulonephritis |
| N03.3 | Chronic nephritic syndrome with diffuse mesangial proliferative glomerulonephritis |
| N03.4 | Chronic nephritic syndrome with diffuse endocapillary proliferative glomerulonephritis |
| N03.5 | Chronic nephritic syndrome with diffuse mesangiocapillary glomerulonephritis |
| N03.6 | Chronic nephritic syndrome with dense deposit disease |
| N03.7 | Chronic nephritic syndrome with diffuse crescentic glomerulonephritis |
| N03.8 | Chronic nephritic syndrome with other morphologic changes |
| N03.9 | Chronic nephritic syndrome with unspecified morphologic changes |
| N04 | Nephrotic syndrome |
| N04.0 | Nephrotic syndrome with minor glomerular abnormality |
| N04.1 | Nephrotic syndrome with focal and segmental glomerular lesions |
| N04.2 | Nephrotic syndrome with diffuse membranous glomerulonephritis |
| N04.3 | Nephrotic syndrome with diffuse mesangial proliferative glomerulonephritis |
| N04.4 | Nephrotic syndrome with diffuse endocapillary proliferative glomerulonephritis |
| N04.5 | Nephrotic syndrome with diffuse mesangiocapillary glomerulonephritis |
| N04.6 | Nephrotic syndrome with dense deposit disease |
| N04.7 | Nephrotic syndrome with diffuse crescentic glomerulonephritis |
| N04.8 | Nephrotic syndrome with other morphologic changes |
| N04.9 | Nephrotic syndrome with unspecified morphologic changes |
| N05 | Unspecified nephritic syndrome |
| N05.0 | Unspecified nephritic syndrome with minor glomerular abnormality |
| N05.1 | Unspecified nephritic syndrome with focal and segmental glomerular lesions |
| N05.2 | Unspecified nephritic syndrome with diffuse membranous glomerulonephritis |
| N05.3 | Unspecified nephritic syndrome with diffuse mesangial proliferative glomerulonephritis |
| N05.4 | Unspecified nephritic syndrome with diffuse endocapillary proliferative glomerulonephritis |
| N05.5 | Unspecified nephritic syndrome with diffuse mesangiocapillary glomerulonephritis |
| N05.6 | Unspecified nephritic syndrome with dense deposit disease |
| N05.7 | Unspecified nephritic syndrome with diffuse crescentic glomerulonephritis |
| N05.8 | Unspecified nephritic syndrome with other morphologic changes |
| N05.9 | Unspecified nephritic syndrome with unspecified morphologic changes |
| N06 | Isolated proteinuria with specified morphological lesion |
| N06.0 | Isolated proteinuria with minor glomerular abnormality |
| N06.1 | Isolated proteinuria with focal and segmental glomerular lesions |
| N06.2 | Isolated proteinuria with diffuse membranous glomerulonephritis |
| N06.3 | Isolated proteinuria with diffuse mesangial proliferative glomerulonephritis |
| N06.4 | Isolated proteinuria with diffuse endocapillary proliferative glomerulonephritis |
| N06.5 | Isolated proteinuria with diffuse mesangiocapillary glomerulonephritis |
| N06.6 | Isolated proteinuria with dense deposit disease |
| N06.7 | Isolated proteinuria with diffuse crescentic glomerulonephritis |
| N06.8 | Isolated proteinuria with other morphologic lesion |
| N06.9 | Isolated proteinuria with unspecified morphologic lesion |
| N07 | Hereditary nephropathy, not elsewhere classified |
| N07.0 | Hereditary nephropathy, not elsewhere classified with minor glomerular abnormality |
| N07.1 | Hereditary nephropathy, not elsewhere classified with focal and segmental glomerular lesions |
| N07.2 | Hereditary nephropathy, not elsewhere classified with diffuse membranous glomerulonephritis |
| N07.3 | Hereditary nephropathy, not elsewhere classified with diffuse mesangial proliferative glomerulonephritis |
| N07.4 | Hereditary nephropathy, not elsewhere classified with diffuse endocapillary proliferative glomerulonephritis |
| N07.5 | Hereditary nephropathy, not elsewhere classified with diffuse mesangiocapillary glomerulonephritis |
| N07.6 | Hereditary nephropathy, not elsewhere classified with dense deposit disease |
| N07.7 | Hereditary nephropathy, not elsewhere classified with diffuse crescentic glomerulonephritis |
| N07.8 | Hereditary nephropathy, not elsewhere classified with other morphologic lesions |
| N07.9 | Hereditary nephropathy, not elsewhere classified with unspecified morphologic lesions |
| E08.2 | Diabetes mellitus due to underlying condition with kidney complications |
| E08.21 | Diabetes mellitus due to underlying condition with diabetic nephropathy |
| E08.22 | Diabetes mellitus due to underlying condition with diabetic chronic kidney disease |
| E08.29 | Diabetes mellitus due to underlying condition with other diabetic kidney complication |
| E09.2 | Drug or chemical induced diabetes mellitus with kidney complications |
| E09.21 | Drug or chemical induced diabetes mellitus with diabetic nephropathy |
| E09.22 | Drug or chemical induced diabetes mellitus with diabetic chronic kidney disease |
| E09.29 | Drug or chemical induced diabetes mellitus with other diabetic kidney complication |
| E10.2 | Type 1 diabetes mellitus with kidney complications |
| E10.21 | Type 1 diabetes mellitus with diabetic nephropathy |
| E10.22 | Type 1 diabetes mellitus with diabetic chronic kidney disease |
| E10.29 | Type 1 diabetes mellitus with other diabetic kidney complication |
| E11.2 | Type 2 diabetes mellitus with kidney complications |
| E11.21 | Type 2 diabetes mellitus with diabetic nephropathy |
| E11.22 | Type 2 diabetes mellitus with diabetic chronic kidney disease |
| E11.29 | Type 2 diabetes mellitus with other diabetic kidney complication |
| E13.2 | Other specified diabetes mellitus with kidney complications |
| E13.21 | Other specified diabetes mellitus with diabetic nephropathy |
| E13.22 | Other specified diabetes mellitus with diabetic chronic kidney disease |
| E13.29 | Other specified diabetes mellitus with other diabetic kidney complication |
| N25 | Disorders resulting from impaired renal tubular function |
| N25.8 | Other disorders resulting from impaired renal tubular function |
| N25.81 | Secondary hyperparathyroidism of renal origin |
| N25.89 | Other disorders resulting from impaired renal tubular function |
| N25.9 | Disorder resulting from impaired renal tubular function, unspecified |
| Q61 | Cystic kidney disease |
| Q61.2 | Polycystic kidney, adult type |
| Q61.3 | Polycystic kidney, unspecified |
| Q61.4 | Renal dysplasia |
| Q61.5 | Medullary cystic kidney |
| Q61.8 | Other cystic kidney diseases |
| Q61.9 | Cystic kidney disease, unspecified |
| N18.1 | Chronic kidney disease, stage 1 |
| N18.2 | Chronic kidney disease, stage 2 (mild) |
| N18.3 | Chronic kidney disease, stage 3 (moderate) |
| N18.4 | Chronic kidney disease, stage 4 (severe) |
| N18.5 | Chronic kidney disease, stage 5 |
| N18.9 | Chronic kidney disease, unspecified |
| N28 | Other disorders of kidney and ureter, not elsewhere classified |
| N28.0 | Ischemia and infarction of kidney |

**Table S3.** Administrative codes for kidney transplant

| ICD Code | Explanation |
|---|---|
| **ICD-9-CM Diagnosis** | |
| V42.0 | Kidney replaced by transplant |
| 996.81 | Complications of transplanted kidney |
| **ICD-9-CM Procedure** | |
| 55.6 | Transplant of Kidney |
| 55.61 | Renal autotransplantation |
| 55.69 | Other kidney transplantation |
| **ICD-10-CM Diagnosis** | |
| Z94.0 | Kidney transplant status |
| T86.10 | Unspecified complication of kidney transplant |
| T86.11 | Kidney transplant rejection |
| T86.12 | Kidney transplant failure |
| T86.13 | Kidney transplant infection |
| T86.19 | Other complication of kidney transplant |
| **ICD-10-PCS Procedure** | |
| 0TS00ZZ | Reposition Right Kidney, Open Approach |
| 0TS10ZZ | Reposition Left Kidney, Open Approach |
| 0TY00Z0 | Transplantation of Right Kidney, Allogeneic, Open Approach |
| 0TY00Z1 | Transplantation of Right Kidney, Syngeneic, Open Approach |
| 0TY10Z0 | Transplantation of Left Kidney, Allogeneic, Open Approach |
| 0TY10Z1 | Transplantation of Left Kidney, Syngeneic, Open Approach |
| **CPT** | |
| 50360 | Renal allotransplantation, implantation of graft; without recipient nephrectomy |
| 50365 | Renal allotransplantation, implantation of graft; with recipient nephrectomy |
| 50380 | Renal autotransplantation, reimplantation of kidney |

Abbreviations. ICD, The International Classification of Diseases; ICD-9-CM, The International Classification of Diseases, Ninth Revision, Clinical Modification; ICD-10-CM, International Classification of Diseases, Tenth Revision, Clinical Modification; ICD-10-PCS, The International Classification of Diseases, 10th Revision, Procedure Coding System; CPT, Current Procedural Terminology

**Table S4.** ICD codes used for history of acute kidney injury

| ICD Code | Explanation |
|---|---|
| **ICD-9-CM Diagnosis** | |
| 584 | Acute kidney failure |
| 584.5 | Acute kidney failure with lesion of tubular necrosis convert |
| 584.6 | Acute kidney failure with lesion of renal cortical necrosis convert |
| 584.7 | Acute kidney failure with lesion of renal medullary [papillary] necrosis |
| 584.8 | Acute kidney failure with other specified pathological lesion in kidney |
| 584.9 | Acute kidney failure, unspecified |
| 593.9 | Unspecified disorder of kidney and ureter (includes renal disease (chronic) not otherwise specified) |
| 997.5 | Urinary complications, not elsewhere classified |
| **ICD-10-CM Diagnosis** | |
| N17 | Acute kidney failure |
| N17.0 | Acute kidney failure with tubular necrosis |
| N17.1 | Acute kidney failure with acute cortical necrosis |
| N17.2 | Acute kidney failure with medullary necrosis |
| N17.8 | Other acute kidney failure |
| N17.9 | Acute kidney failure, unspecified |
| N28.9 | Disorder of kidney and ureter, unspecified |

Abbreviations. ICD, The International Classification of Diseases; ICD-9-CM, The International Classification of Diseases, Ninth Revision, Clinical Modification; ICD-10-CM, International Classification of Diseases, Tenth Revision, Clinical Modification

**Table S5.** Administrative codes used for renal-replacement therapy

| ICD Code | Explanation |
|---|---|
| **ICD-9-CM Diagnosis** | |
| V45.12 | Noncompliance with renal dialysis |
| V56.0 | Encounter for extracorporeal dialysis |
| V56.8 | Encounter for other dialysis |
| V56.1 | Fitting and adjustment of extracorporeal dialysis catheter |
| V56.2 | Fitting and adjustment of peritoneal dialysis catheter |
| V56.32 | Encounter for adequacy testing for peritoneal dialysis |
| V45.1 | Renal dialysis status |
| V45.11 | Renal dialysis status |
| 996.56 | Mechanical complication due to peritoneal dialysis catheter |
| 996.68 | Infection and inflammatory reaction due to peritoneal dialysis catheter |
| 792.5 | Cloudy (hemodialysis) (peritoneal) dialysis effluent |
| **ICD-9-CM Procedure** | |
| 39.95 | Hemodialysis |
| 54.98 | Peritoneal dialysis |
| **ICD-10-CM Diagnosis** | |
| Z91.15 | Patient's noncompliance with renal dialysis |
| Z49.31 | Encounter for adequacy testing for hemodialysis |
| Z49.32 | Encounter for adequacy testing for peritoneal dialysis |
| Z49.01 | Encounter for fitting and adjustment of extracorporeal dialysis catheter |
| Z49.02 | Encounter for fitting and adjustment of peritoneal dialysis catheter |
| Z49.32 | Encounter for adequacy testing for peritoneal dialysis |
| T85.71XA | Infection and inflammatory reaction due to peritoneal dialysis catheter, initial encounter |
| T85.611A | Breakdown (mechanical) of intraperitoneal dialysis catheter, initial encounter |
| T85.621A | Displacement of intraperitoneal dialysis catheter, initial encounter |
| R88.0 | Cloudy (hemodialysis) (peritoneal) dialysis effluent |
| T85.631A | Leakage of intraperitoneal dialysis catheter, initial encounter |
| T85.71XA | Infection and inflammatory reaction due to peritoneal dialysis catheter |
| T85.71XS | Infection and inflammatory reaction due to peritoneal dialysis catheter, sequela |
| Z99.2 | Dependence on renal dialysis |
| T85.71XD | Infection and inflammatory reaction due to peritoneal dialysis catheter, subsequent encounter |
| Z99.2 | Dependence on renal dialysis |
| **ICD-10-PCS Procedure** | |
| 5A1D00Z | Performance of Urinary Filtration, Single |
| 5A1D60Z | Performance of Urinary Filtration, Multiple |
| 3E1M39Z | Irrigation of Peritoneal Cavity using Dialysate, Percutaneous Approach |
| **CPT** | |
| 90935 | Hemodialysis procedure with single evaluation by a physician or other qualified health care professional |
| 90937 | Hemodialysis procedure requiring repeated evaluation(s) with or without substantial revision of dialysis prescription |
| 90945 | Dialysis procedure other than hemodialysis (eg, peritoneal dialysis, hemofiltration, or other continuous renal replacement therapies), with single evaluation by a physician or other qualified health care professional |
| 90947 | Dialysis procedure other than hemodialysis (eg, peritoneal dialysis, hemofiltration, or other continuous renal replacement therapies) requiring repeated evaluations by a physician or other qualified health care professional, with or without substantial revision of dialysis prescription |
| 90999 | Unlisted dialysis procedure, inpatient or outpatient |

Abbreviations. ICD, The International Classification of Diseases; ICD-9-CM, The International Classification of Diseases, Ninth Revision, Clinical Modification; ICD-10-CM, International Classification of Diseases, Tenth Revision, Clinical Modification; ICD-10-PCS, The International Classification of Diseases, 10th Revision, Procedure Coding System; CPT, Current Procedural Terminology

**Table S6.** Data elements that are used to run CKD phenotyping algorithm

| **Features** | **Description** | **Format** |
|---|---|---|
| patient_deiden_id | Deidentified Patient ID | Strings |
| encounter_deiden_id | Deidentified Encounter ID | Strings |
| admit_datetime | Hospital Admission Date and Time | Date and Time |
| dischg_datetime | Discharge Date and Time | Date and Time |
| birth_date | Birth Date | Date and Time |
| sex | Sex | Strings |
| race | Race | Strings |
| ethnicity | Ethnicity | Strings |
| patient_type | Patient Type | Strings |
| start_date | Diagnosis Start Date | Date |
| diag_code | Diagnosis Code | Strings |
| diag_icd_type | Diagnosis Code Type | Strings |
| proc_date | Procedure Date | Date |
| proc_code_type | Procedure Code Type | Strings |
| proc_code | Procedure Code | Strings |
| lab_result | Lab Result | Float |
| lab_unit | Lab Unit | Strings |
| inferred_specimen_datetime | Inferred Specimen Taken Date and Time | Date and Time |
| stamped_and_inferred_loinc_code | Stamped and Inferred LOINC Code | Strings |

**Table S7**. Data elements that are used to run AKI phenotyping algorithm

| **Features** | **Description** | **Format** |
|---|---|---|
| patient_deiden_id | Deidentified Patient ID | Strings |
| encounter_deiden_id | Deidentified Encounter ID | Strings |
| Code | CPT code | Strings |
| Proc Date | Date for CPT code | Date |
| vital_sign_measure_name[a] | The type of vital signs measured | Strings |
| meas_value[a] | The measured value of the vital | Strings |
| recorded_time[a] | The date and time of measurement | Date and Time |
| hemodialysis_intake[a] | Intake value for hemodialysis | Float |
| hemodialysis_output[a] | Output value for hemodialysis | Float |
| peritoneal_dialysis_intake[a] | Intake value for Peritoneal Dialysis | Float |
| peritoneal_dialysis_output[a] | Output value for Peritoneal Dialysis | Float |
| observation_datetime[a] | Date and time for measurement for dialysis intake and output | Date and Time |
| lab_result | Lab Result | Float |
| lab_unit | Lab Unit | Strings |
| inferred_specimen_datetime | Inferred Specimen Taken Date and Time | Date and Time |
| stamped_and_inferred_loinc_code | Stamped and Inferred LOINC Code | Strings |

[a] Data elements that are used for determining dialysis status in addition to CPT codes. Patient is considered to be on dialysis when the type of vital signs measured is "Treatment Type" and the measured value of the vital is *CVVH* or *CVVHD* or *CVVHDF or if there is non-zero hemodialysis or peritoneal dialysis intake or output.*

Abbreviations. CVVH, Continuous Veno-Venous Hemofiltration; *CVVHD, Continuous Venovenous Hemodialysis; CVVHDF, Continuous Venovenous Hemodiafiltration*